\documentclass[12pt]{article}
\def\be{\begin{equation}}
\def\ee{\end{equation}}

\def\H{\mathrm{H}}
\def\T{\mathrm{T}}
\def\t{\mathrm{t}}
\def\q{\mathrm{q}}
\def\g{\mathrm{g}}
\def\R{\mathrm{R}}
\def\ApJ{\sl Astrophys. J.}
\def\MNRAS{\sl Mon. Not. R. Astron. Soc.}

\usepackage{amsmath}
\usepackage{psfig,epsfig}
\usepackage{latexsym}
\usepackage{pifont}

\title{\bf 
{Age of High Redshift Objects - a Litmus Test for the Dark Energy Models}}

\author{Deepak Jain(1)\footnote{E--mail :
deepak@physics.du.ac.in} and Abha Dev(2)\\
	{1. Deen Dayal Upadhyaya College, University of Delhi, Delhi 110015, India}\\
        { 2. Miranda House, University of Delhi, Delhi 110007, India}
	}

\begin {document}

\maketitle

\begin{center}
\Large{\bf Abstract}
\end{center}
The discovery of the quasar, the APM $08279+5255$ at z
= 3.91 whose age is \\ 2-3 Gyr has once again led to ``age crisis''.
The noticeable fact about this object
is that it cannot be accommodated in a universe with $\Omega_m = 0.27$,
currently accepted value of matter density
parameter and $\omega = \mathrm{constant}$. In this work, we
explore the concordance of various dark energy parameterizations
($w(z)$ models) with
the age estimates of the old high redshift objects. It is alarming to
note that the quasar cannot be accommodated in any dark energy
model even for  $\Omega_m = 0.23$, which corresponds to $1 \sigma$
deviation below the best fit value provided by WMAP. There is a
need to look for alternative cosmologies or some other dark energy
parameterizations which allow the existence of the high redshift 
objects.

\large
\baselineskip=20pt

   \vfill
      \eject

\baselineskip = 20pt

\begin {section} {Introduction}

\noindent A lesson given by the history of cosmology is that the concept of the
cosmological constant revives in the days of crisis. In the recent past,
perhaps the most important reason to reconsider the cosmological
constant was to resolve the age crisis in which the universe 
become younger than its constituents! For example, 
the discoveries of a 3.5 Gyr old galaxy at
z = 1.55 and of a 4.0 Gyr old galaxy at z = 1.43 have been proved to be
incompatible in Einstein-de Sitter universe \cite{dun,spin,dun1}.

\vskip 0.5 cm
\noindent  It was shown that a flat FRW type cosmological model 
dominated by self-interacting, unclustered fluid with  negative
pressure collectively known as ``dark energy'' can accommodate these
galaxies and hence alleviate the ``age problem'' \cite{lima}. This idea of
dark energy is strongly supported by the 
observational results of the anisotropy of cosmic microwave background  
radiation, distance measurements of Type Ia Supernovae and Sloan
Digital Sky Survey \cite{obs}. One simple candidate of this dark energy is the 
cosmological constant parametrized by equation of state $ \mathrm{p} = \omega
\, \rho$ with $ \omega = -1$. 
Although the cold dark matter cosmological constant models are
consistent with the present observations, however, there are some
fine tuning problems with the unevolving cosmological constant \cite{cc}.
 
\vskip 0.5cm
\noindent This has led theorists to explore various dark energy
models \cite{de}. However, the nature of dark energy is still unknown. It is not     
plausible to check every dark energy model by using the observational  
data. Therefore, model independent probe of dark energy is a good
alternative to study its nature. The most common way of measuring 
dark energy properties has been to parameterize the $\omega$ by one or
two free parameters and  constraining these by fitting the observed
data. In the present work we focus our attention to four different
popular parameterizations of dark energy. In these models the dark
energy components evolves with time. These models are most
commonly used in the literature to study the properties of dark  
energy. 
\vskip 0.5cm

\noindent 
In this paper we discuss  observational constraints on
various parameterizations of dark energy from  age  measurements of
old high redshift galaxies (OHRG). These are, namely, the LBDS 53W069, a 4
Gyr-old radio galaxy at z = 1.43 and  the  LBDS 53W091, a 3.5
Gyr-old radio galaxy at z = 1.55 .

\vskip 0.5cm
\noindent 
The discovery of a quasar, the APM $08279+5255$ at z
= 3.91 whose age is 2-3 Gyr has once again led to ``age crisis'' \cite{q}.
The noticeable fact about this object
is that it cannot be accommodated in a universe with $\Omega_m = 0.27$,
currently accepted value of matter density
parameter and $\omega = \mathrm{constant}$ \cite{al}. It is shown that
other dark energy 
scenarios  i.e. brane model, generalized  Chaplygin gas model etc.,
are also not compatible with the existence of this object unless we
lower the value of $\mathrm{H}_0$ and $\Omega_m$ \cite{al1}. Therefore 
in the coming years this object may  act as `litmus test ' for every
new dark energy scenario.

\vskip 0.5cm
\noindent {\it The Primary objective of this paper is to use age estimates
of the old high redshifts objects to constrain the cosmological
models in which dark energy equation of state parameter evolves with time}.

\noindent The paper is organised as follows: In Section 2 we
introduce the various dark energy parameterizations. Cosmic-age
redshift is described in the Section 3. Section 4 discusses the
results and summarizes our conclusions.
\end {section}

\begin{section}{$\omega(z)$ Models}
\noindent We consider spatially flat, homogeneous and isotropic cosmologies.
In the presence of non relativistic matter and
a dark energy component, the Hubble constant varies with redshift as:
 
\be
 \H^2(z)\, \equiv \,({\dot{\R}\over \R})^{2}\; = \;\H_{0}^{2}\left[\Omega_m
 \, ({\R_0\over \R})^{3} +
 \Omega_x\, f\right ]
 \ee
 
 \noindent where the dot represents derivative with respect to time. Further
 $$ \Omega_m = {8 \;\pi\; \mathrm{G} \over 3 \H_0^{2}}\rho_{m0}\;\;, \;\Omega_x =
 {8 \;\pi\; \mathrm{G} \over 3 \H_0^{2}}\rho_{x0} $$ where $\H_0$ is the
 Hubble
 constant at the present epoch, while $\rho_{m0} \;\;{\rm{and}}
\;\;\rho_{x0} $ are the 
 non relativistic matter
 density and the dark energy density respectively at the present
epoch. Also ${f} = \rho_x/{\rho_x}_o$.

\vskip 0.5cm
\noindent
In this paper, we work with the following specific parameterizations for the
variation of $\omega$ with redshift, i.e.,

\begin{equation}
\mathrm{P0}:\quad \quad  \omega(z) = \omega_o,
\end{equation}
\begin{equation}
\mathrm{P1}:\quad \quad \omega(z) = \omega_o + \omega_1 z,
\end{equation}
\begin{equation}
\mathrm{P2}: \quad \quad \omega(z) = \omega_o - \omega_2\,{\rm{ln}}(1 + z),
\end{equation}
\begin{equation}
\mathrm{P3}: \quad \quad \omega(z) = \omega_o +  \omega_3\left(\frac{z}{1 + z}\right),
\end{equation}
and
\begin{equation}
\mathrm{P4}:\quad \quad \omega(z) = \omega_o +  \omega_4\,\frac{z}{(1 + z)^2}.
\end{equation}

\noindent Here $\omega_o$ is the current value of the
equation-of-state parameter and $\omega_j$ 
($j = 1, 2, 3,4$) are free parameters quantifying the time-dependence of the dark energy
and these are to be constrained using the observational data \cite{jain}. Note
that the equation of state 
of the cosmological constant can be always recovered by taking $\omega_j = 0$ and
$\omega_o = -1$. 

\vskip 0.5cm
\noindent
The model P0 is a simple generalization of dark energy with constant equation
of state parameter. This ansatz is referred to as `quintessence' ($\omega_o
\le 1$) in the literature.
Constraints on the Taylor expansion (P1 model) were firstly studied 
by Cooray \& Huterer \cite{huterer} by using SNe Ia data, gravitational
lensing statistics and globular clusters ages. As
commented in Ref. \cite{huterer}, 
P1 is a good approximation for most quintessence models out to
redshift of a few and it 
is exact for models where the equation of state is a constant or
changing slowly. P1, 
however, has serious problems in explaining age estimates of high-$z$
objects since it 
predicts very small ages at $z \geq 3$ \cite{al1}. In reality, P1
blows up at high-redshifts as $e^{3\omega_1z}$ for values of $\omega_1
> 0$. The
empirical fit P2 was introduced by Efstathiou \cite{efs} who argued that for a wide
class of potentials associated to dynamical scalar field models the evolution of
$\omega(z)$ at $z \le 4$ is well approximated by equation (2). 
P3 was recently proposed by Linder \cite{linder} (see also \cite{pad}) 
aiming at solving undesirable
behaviors of P1 at high redshifts. According to \cite{krat}, 
this parametrization is a good fit for many theoretically conceivable 
scalar field potentials, as well as for
small recent deviations from a pure cosmological constant behavior ($\omega = -1$) (see
also \cite{wsps,teg,padn}) for other parameterizations] but this
parametrization blows up exponentially in the future  as $ R
\rightarrow \infty$ for $\omega_3 > 0$. In P4 parametrization, the dark
energy component has the same equation of state at the present epoch
(z = 0) and at $ z\gg 1$ with rapid variation at low z \cite{padn1}.  

\vskip 0.5cm
\noindent
Since equations (2-6) represent separately conserved components, 
it is straightforward to
show that the ratio ${f} = \rho_x/{\rho_x}_o$ for (P0)-(P4) 
evolves, respectively, as
\begin{equation}
\mathrm{P0}:\quad \quad f_0 = (\frac{R_o}{R})^{3(1 + \omega_o)},
\end{equation}
\begin{equation}
\mathrm{P1}:\quad \quad f_1 = (\frac{R_o}{R})^{3(1 + 
\omega_o - \omega_1)}\rm{exp}\left[3 
\omega_1 (\frac{R_o}{R} - 1)\right],
\end{equation}
\begin{equation}
\mathrm{P2}:\quad \quad  f_2 = (\frac{R_o}{R})^{3\left[1 + \omega_o
-\frac{\omega_2}{2}\rm{ln}(\frac{R_o}{R})\right]},
\end{equation}
\begin{equation}
\mathrm{P3}:\quad \quad f_3 = (\frac{R_o}{R})^{3(1 + \omega_o +
\omega_3)}\rm{exp}\left[3\omega_3(\frac{R}{R_o} - 1)\right],
\end{equation}
\noindent and
\begin{equation}
\mathrm{P4}:\quad \quad f_4 = (\frac{R_o}{R})^{3(1 + \omega_o)}\rm{exp}\left[\frac{3}{2} 
\omega_4 (\frac{R_o}{R} - 1)\right].
\end{equation}
\noindent Here the subscript $o$ denotes present day quantities and 
$\R(t)$ is the cosmological
scale factor. The age of the universe at the redshift z is given by
 \begin{eqnarray}
 {\H}_0\; t(z) &= &{\H}_0\;\int_z^{\infty}{dz'\over(1 + z'){\H}(z')}
 \nonumber \\
 & & \nonumber \\
 && =  \int_z^{\infty}{dz'\over(1 + z')\sqrt{\Omega_m(1 + z')^3 +
 \Omega_x f}}. 
 \end{eqnarray}

\end{section}

\begin{section}{Cosmic Age-Redshift Test }

\noindent
In order to constrain the dark energy models under consideration
 from the age estimate of the above
mentioned quasar we follow the line of thought as given in
ref. \cite{lima}. Firstly, the age of the universe at a 
given redshift has to be  greater than or at least equal to the age 
of its oldest objects 
at that redshift. Hence this test provides lower bound on $\omega_0$ 
and $\omega_j$. This 
can be checked if we define the dimensionless ratio:
\be
{t(z)\over t_{\rm obj}} = {{\H}_0\, t(z)\over {\T}_{\rm obj} = {\H}_0 \,t_{\rm obj}} \ge 1\,\,.
\ee
\noindent Here $t_{\rm obj}$ is the age of an old object at a
given redshift. For every
high redshift object, ${\T}_{\rm obj} = {\H}_0 \,t_{\rm obj}$ is a dimensionless age parameter. 
The error bar
on ${\H}_{0}$ determines the extreme value of $\T_{\rm obj}$. The lower  
limit on ${\H}_0$ is updated to nearly $10\%$ of accuracy by Freedman \cite{Freedman}: 
$\H_0 = 72 \pm 8$ km/sec/Mpc. We use minimal value of the
 Hubble constant, $ {\H}_0= 64$ km/sec/Mpc, to
get strong conservative limit.

Komossa and Hasinger (2002) \cite{KH} have estimated the age of the quasar
APM $08279+5255$ at redshift z= 3.91 to be between the interval 2-3 Gyr. 
They use an Fe/O =3 abundance ratio (normalised to the solar value) 
as inferred from X-ray observations to
draw the conclusion. An age of 3 Gyr is concluded from the 
temporal evolution of Fe/O ratio in the giant elliptical model (M4a) 
of Hamman and Ferland \cite{HF}. The age estimate of 2 Gyr is inferred by using
the ``extreme model'' M6a of Hamman and  Ferland for which the Fe/O
evolution is faster 
and Fe/O = 3 is reached after 2 Gyr. Friaca, Alcaniz and Lima (2005) \cite{al1}
reevaluate the age  
for APM 0879+5255 by using a chemodynamical model for the evolution of 
spheroids. They quote the age of this old quasar at z = 3.91 as 2.1 Gyr.  
To assure the robustness of our analysis 
we use the lower age estimate for our calculations.
The 2 Gyr old quasar at z = 3.91 gives ${\T}_{\q} = 2.0 \H_0$ Gyr and 
hence $0.131 \le {\T}_{\q} \le 0.163$. It thus follows that  ${\T}_{\q} \ge
0.131$. 

Similarly, for LBDS 53W069, a 4
Gyr-old radio galaxy at z = 1.43, we have ${\T}_{\g} \ge 0.261$ and for
the  LBDS 53W091, a 3.5 Gyr-old radio galaxy at z = 1.55, ${\T}_{\g} \ge 0.229$.

\end{section}

\begin{section}{Results and Discussions}

\noindent It is now a well accepted fact that the universe is
accelerating and is dominated by a smoothly distributed cosmic fluid, 
referred to as 'dark energy'. A number of observational tests have been
proposed in the literature to study the nature of dark energy,
including SNe Type Ia luminosity distances, gravitational lensing, 
Cosmic Microwave Background anisotropy, angular size-redshift
relationship of compact
radio sources, Lyman-alpha forest  etc.
Essentially most  of the proposed observational tests are based on the  
measurement of distance-redshift relationship. The age estimate of old
high redshift objects provides an independent way of studying various
dark energy models. As opposed to the other tests listed above, this
method is based on the time-dependent observables. In this work, we
explore the concordance of various dark energy parameterizations with
the age estimates of the old high redshift objects.
\vskip 0.5 cm
\noindent
Fig.\,1 summarizes the results for the P0 model. The model is a simple
generalization of dark energy with constant equation 
of state parameter. This parametrization fails to accommodate the
old quasar for $\Omega_m = 0.27$ as shown in Fig.\,1 . This is a
serious problem. The two old galaxies put a upper bound  
on $\omega_0$ as: $\omega_0 \leq -0.32$ for LBDS 53W069, a 4
Gyr-old radio galaxy at z = 1.43 to exist and $\omega_0 \leq -0.24$ for
the  LBDS 53W091, a 3.5 Gyr-old radio galaxy at z = 1.55 to exist. In
case the dark energy is  modeled as 
cosmological constant  ($\omega_0 = -1$), the old quasar is accommodated
only if $\Omega_m < 0.21$.
\vskip 0.5 cm
\noindent
The results for P1 model are shown in Fig.\,2. For a given old galaxy,
the line represents the minimal value of its age parameter (${\T}_{\g} = {\H}_0t_{\g}$). 
Of the two galaxies, 
the radio galaxy, LBDS 53W069, at z = 1.43 provides tighter constraints.
For this galaxy, ${\T}_{\g} \ge
0.261$. The age estimates provide the constraints: $\omega_0 \le -0.31$
and $\omega_1 \le  0.96$ for $\Omega_m = 0.27$. The constraints change to
$\omega_0 \le -0.28$ and $\omega_1 \le  1.08$ for $\Omega_m =
0.23$. Here, we have worked with the following ranges: 
$-2 \le \omega_0 \le 0$ and $ 0 \le \omega_1 \le 4.0$. 
We once again find that the parameterization fails to accommodate the
old quasar for $\Omega_m = 0.27$. It fails to do so even for $\Omega_m = 0.23$.    
\vskip 0.5 cm
\noindent
In Fig.\,3, we display the results for P2 model. For this model, we 
work with the following ranges: $-2 \le \omega_0 \le 0$ and $ -2 \le \omega_2 \le 0$. 
The radio galaxy LBDS 53W069 puts the constraints: $\omega_0 \le -0.31$
and $\omega_2 \ge  -1.95$ for $\Omega_m = 0.27$. The constraints change to 
$\omega_0 \le 0.-0.28$ and $\omega_2 \ge  -1.99$ for $\Omega_m = 0.23$. 
The parameterization fails to accommodate the quasar even for $\Omega_m = 0.23$. 

\vskip 0.5 cm
\noindent In Fig.\,4 we show the parametric space $ \omega_0 -
\omega_3 $ for the P3 model. We 
work with the ranges: $-2 \le \omega_0 \le 0$ and $ 0 \le \omega_2 \le 4.0$. 
The tighter constraint is given by 4.0-Gyr
galaxy(53W069) at z = 1.43 as expected. The constraints are: 
$\omega_0 \le -0.31$ and $\omega_3 \le  3.29$ for $\Omega_m = 0.27$. 
The constraints change to $\omega_0 \le 0.-0.28$ and $\omega_3 \le 3.39$ 
for $\Omega_m = 0.23$.

\vskip 0.5 cm
\noindent Fig.\,5 shows the parametric space $ \omega_0 - \omega_4 $ for P4
model. The ranges for the parameters are: $-2 \le \omega_0 \le 0$ and
$ 0 \le \omega_4 \le 4.0$. For $\Omega_m =0.27$, we get the constraint 
$\omega_0 \le -0.31$ while the entire range of $\omega_4 $
is allowed. For $\Omega_m = 0.23$, we get $\omega_0 \le -0.28$.
  
\vskip 0.5 cm
\noindent
It is interesting to note the following points:

\noindent 1. If we concentrate on the two galaxies, for all the models
considered above, galaxy data rules out non-accelerating models of
the universe. The main issue whether the dark  energy is a cosmological
constant or if it is evolving with redshift, cannot be resolved with
these data points. Since  $\omega_0 < -0.31$ and wide range of $
\omega_j$ ( j = 1,2,3,4)  is allowed for $\Omega_m = 0.27$. Therefore,
with these data points the age-redshift test does not rule out
cosmological  
constant as a dark energy candidate.
\vskip 0.3cm
\noindent 2. {\it The old quasar cannot be accommodated in any of the
 dark energy parameterizations} for any range of $w_0$ and $w_j$ even for  
$\Omega_m = 0.23$ which corresponds to $1 \sigma$
deviation below the best fit value provided by WMAP.

\vskip 0.5 cm

\noindent 
The quasar can be accomodated if the values of ${\H}_0$ or/and 
$\Omega_m$ are further lowered down from their
currently accepted values. However, reduction of the matter contributions 
would disturb the galaxy formation.

\vskip 0.5cm
\noindent
Since almost all the age of the universe is at low redshift 
(z = 0-2), the galaxies may have formed nearly at the same
epoch, regardless of their constraints on the redshift space. The
effect of dark energy on the galaxy formation epoch has been studied by
Alcaniz and Lima (2001)\cite{a}. A large matter contribution results
in a larger value of $z_f$. It is shown that lower the value of $w$, lower the
value of redshift of formation of galaxies, $z_f$. For example, the galaxy
53W091 with $\Omega_m$ = 0.3, $ w = -1/3$ gives $z_f \ge 20.3$. The
$z_f \ge 5.2$ for the same galaxy with $w = -1$. P. J. McCarthy et
al. (2004) report a $z_f = 2.4$ from conservative age estimates for 20
colour selected red galaxies with $1.3 < z < 2.2$ \cite{mc}. This may require
$w <-1$ with $\Omega_m \sim 0.27$.  

\vskip 0.5 cm 

\noindent Recently, this old quasar APM $08279+5255$ is studied in detail 
in reference with various other dark energy  
scenarios \cite{al,al1}. Friaca, Alcaniz and Lima also work with the
P0, P1 and P3 parameterizations using the old quasar \cite{al1}. They, however,
consider $w_o, w_1 $and $w_3$ as fixed parameters.
They report that this object is not compatible with any
model unless the values of ${\H}_0$ or/and $\Omega_m$ are further
lowered down from their current accepted values.

\vskip 0.5 cm
\noindent As discussed in Section 3, the estimated age of the quasar APM
$08279+5255$ lies between 2-3 Gyr.   
In this paper, we use the lower age estimate of 2 Gyr for the quasar
in order to increase the robustness of our result. 

\vskip 0.5 cm

\noindent
It looks like we are once again in a situation of crisis. There is a
need to look for alternative cosmologies or some other dark energy
parameterizations which allow the existence of the high redshift 
objects. Linear coasting cosmology \cite{get} and Brane world models
 \cite{v} accommodates this old 
quasar comfortably. These models are also concordant with the host of
other cosmological observations \cite{v,linear}.

\vskip 0.5 cm 
\noindent In this paper we use the method based on the absolute
age determination of the high redshift objects. A quite similar
approach is also used by Capozziello et al. \cite{cap}.
They  use the look back time to high
redshift objects (clusters of galaxies) rather than their cosmic age to
constrain various dark energy models. They, however, point out that
galaxy clusters are not 
good candidates for such a study because it is difficult to detect a
enough number of member galaxies at redshift greater than
1.3. Similarly, cosmic age from globular 
clusters is used as a tool to constrain  dark energy
parameterizations \cite{feng}. The results from this method show that
age limits plays significant role in understanding the properties of dark
energy. 

\vskip 0.5cm
\noindent Jimenez and Loeb propose the use of relative
ages as a tool for constraining the cosmological
parameters \cite{jim}. They emphasize that the relative age is better
determined than the absolute age because systematic effects on the
absolute scale are factored out( for a fractional age difference $<<$
1). The statistical significance of the results depends upon the
samples of passively evolving galaxies with high-quality
spectroscopy. All samples need to have similar metallicities and low
star formation rates \cite{sim}. But the
small number of galaxies in the sample again donot give enough
accuracy to the constrain $w(z)$. 

\vskip 0.5 cm
\noindent The calibration of the absolute age of high  redshift
objects is subject to
observational (precise measurements of their distances, turn off
luminosity) and theoretical (all the aspects of stellar astronomy)
uncertainties \cite{st}. Never-the-less it is an independent elementary
means  
to test cosmological models. In future, with more input
data, this simple tool may turn into a powerful one.

\end{section}

\begin{section}*{Acknowledgments}
The authors are thankful to J. S. Alcaniz and  Boud Roukema for useful
suggestions. 
The authors wish to extend their thanks to the referee for  valuable
comments.
\end {section}

\begin{figure}[ht]
\centerline{
\epsfig{figure=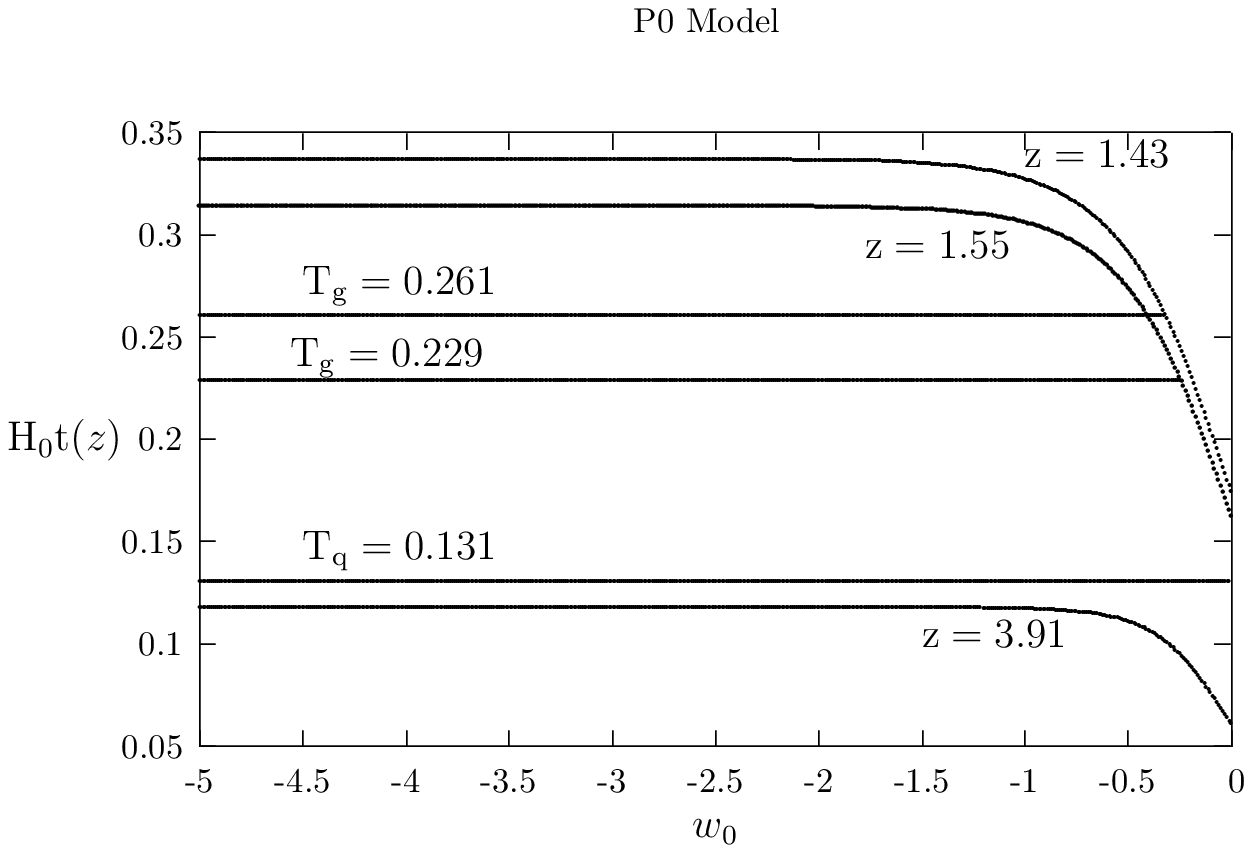,width=1.5\textwidth}}
\vspace{-5.2in} 
\caption{${\H}_0{\t}(z)$ as a function of $\omega_0$.} 
\end{figure}
\begin{figure}[ht]
\centerline{
\epsfig{figure=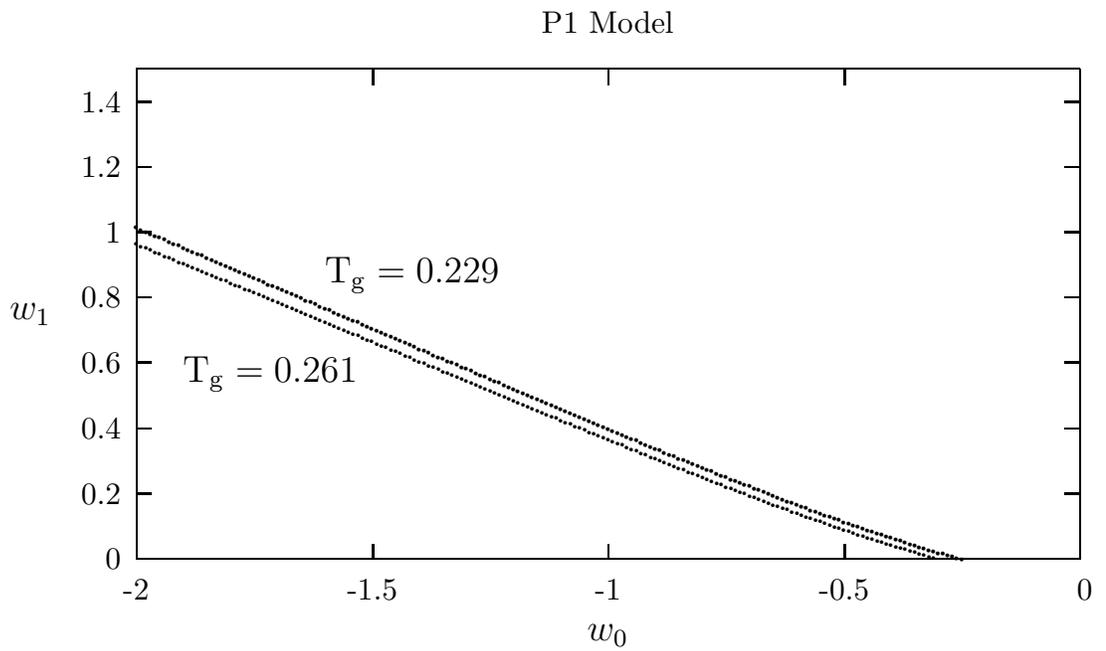,width=1.5\textwidth}}
\vspace{-5.2in} 
\caption{The two contours in the parametric space correspond to 
the constant values of ${\H}_0{\t}_{\rm obj}$
for the two galaxies.The space below the curve ${\T}_{\g} = 0.261$ 
is the allowed region.} 

\end{figure}
\begin{figure}[ht]
\centerline{
\epsfig{figure=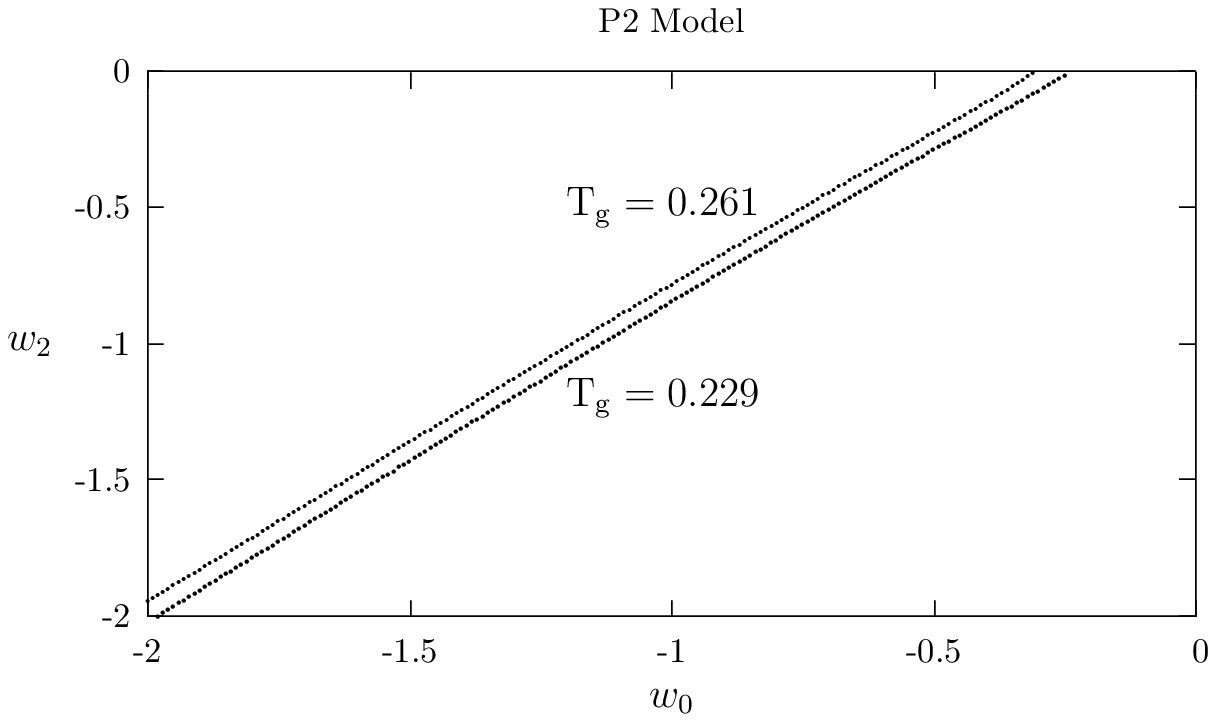,width=1.5\textwidth}}
\vspace{-5.2in} 
\caption{The space above the curve ${\T}_{\g} = 0.261$ gives the possible combinations of
($\omega_0,\omega_2$) which allow both the old high redshift galaxies to exist.} 

\end{figure}
\begin{figure}[ht]
\centerline{
\epsfig{figure=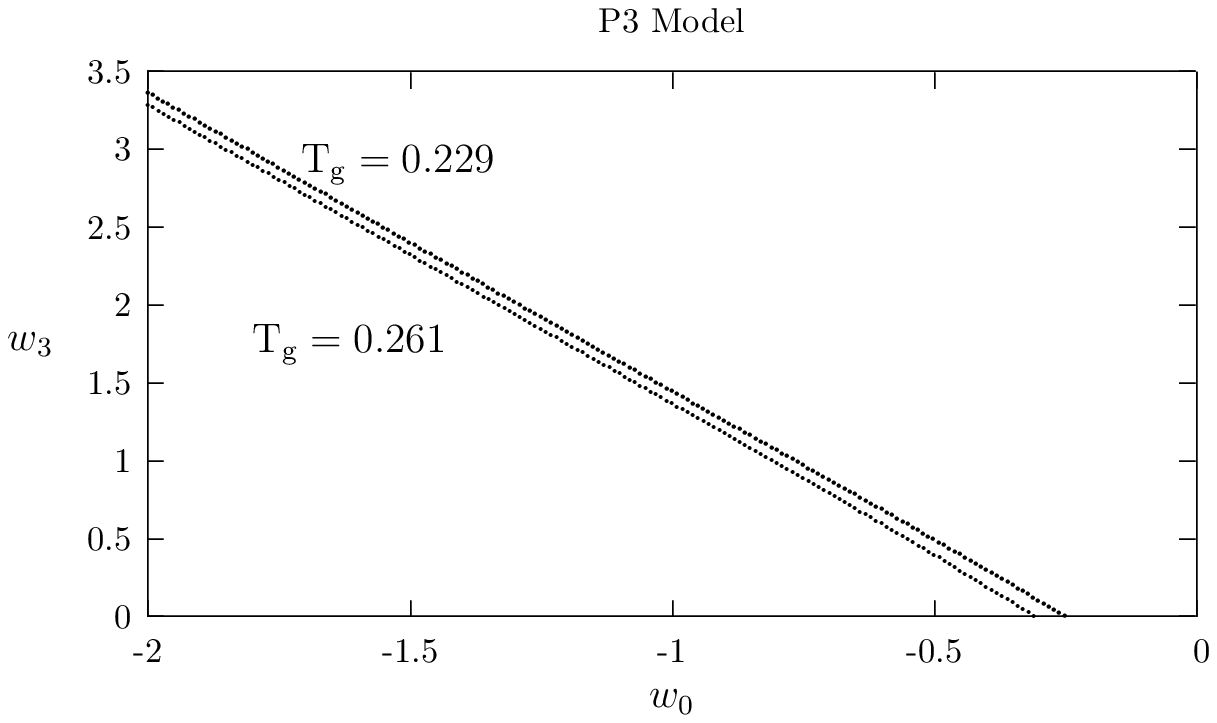,width=1.5\textwidth}}
\vspace{-5.2in} 
\caption{The space below the curve ${\T}_{\g} = 0.261$ gives the possible combinations of
($\omega_0,\omega_3$) which allow both the old high redshift galaxies to exist.} 

\end{figure}
\begin{figure}[ht]
\centerline{
\epsfig{figure=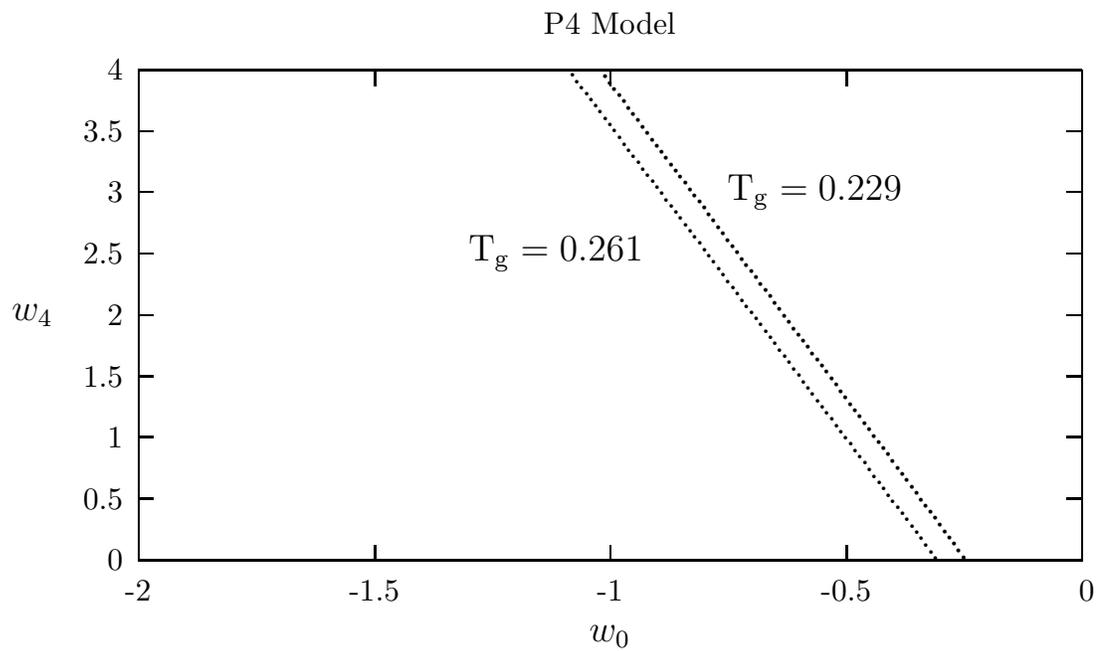,width=1.5\textwidth}}
\vspace{-5.2in} 
\caption{The region left to the ${\T}_{\g} = 0.261$ gives the possible combinations 
($\omega_0, \omega_4$) for which both the galaxies can exist.} 

\end{figure}

\begin{table}
\begin{center}
\begin{tabular}{|l|l|l|}\hline\hline
$Model$ & Constraints from the Galaxies & Constraints from the Quasar   \\   \hline
\hline
&&\\
 P0 & $\omega_0 \leq -0.32$  & Not Accommodated   \\
 P1 & $\omega_0 \le -0.31$, $\omega_1 \le  0.96$   & Not accommodated \\
 P2 &$\omega_0 \le -0.31$, $\omega_2 \ge  -1.95$   & Not accommodated \\
P3 &$\omega_0 \le -0.31$, $\omega_3 \le  3.29$ & Not accommodated \\
P4 & $\omega_0 \le -0.31$, entire range of $\omega_4 $
is allowed  & Not accommodated \\

\hline
\end{tabular}
\caption{The constraints on $\omega_0$ and $\omega_j$ are with
 $\Omega_m = 0.27$ and
$ {\H}_0= 64$ km/sec/Mpc. } 
\end{center}
\end{table}


\begin {thebibliography}{99}

\bibitem{dun} J. S. Dunlop et al., Nature, {\bf 381}, 581 (1996)

\bibitem{spin} S. Spinrard et al., {\ApJ}, {\bf 484}, 581 (1997)

\bibitem{dun1} J. S. Dunlop, in The Most Distant Radio Galaxies,
  ed. H. J. A. Rottgering, P. Best \& M.D.Lehnert, Dordrecht:Kluwer, 71
  (1998).

\bibitem{lima}J. S. Alcaniz  and J. A. S. Lima , {\ApJ}, {\bf 521}.
  L87 (1999);
J. A. S. Lima and  J. S. Alcaniz, {\MNRAS}, {\bf 317}, 893 (2000).

\bibitem{obs}
A. Riess et al., {\it Astron. J.}, {\bf 116}, 1009 (1998);
S. Perlmutter et al., {\ApJ}, {\bf 517}, 565 (1999); D. N. Spergel  et
al.,{\ApJ   Suppl.}, {\bf 148}, 175 (2003);
M. Tegmark  et al., {\it Phys. Rev. D}, {\bf 69}, 103501
(2004); A. Riess et al., {\ApJ},{\bf 607}, 665 (2004).


\bibitem{cc}V. Sahni and A. A. Starobinsky, {\it
  Int. J. Mod. Phys. D}, {\bf 9}, 373 (2000); T. Padmanabhan,{\it
  Phys. Rep.}{\bf 380}, 235 (2003); P. J.Peebles and B. Ratra, {\it
  Rev. Mod. Phys.} {\bf 75}, 559 (2003); T. Padmanabhan, {\bf
  astro-ph/0411044}. 

 \bibitem{de} 
J. A. S. Lima, {\it Braz. Jr. Phys}, {\bf 34}, 194 (2004)
U. Alam, V. Sahni and A. A. Starobinsky {\bf astro-ph/0403687};
V. Sahni, {\bf astro-ph/0403324}; 
T. Padmanabhan,  {\bf gr-qc/0503107} .
\bibitem{q}
G. Hasinger et al., {\ApJ}, {\bf 573}, L77 (2002); S. Komossa and
G. Hasinger, {\bf astro-ph/0207321}.

\bibitem{al} J. S. Alcaniz, J. A. S. lima and J. V. Cunha, {\MNRAS}, {\bf
  340}, L39 (2003).


\bibitem{al1} A. C. S. Friaca, J. S. Alcaniz and J. A. S. lima , {\bf
  astro-ph/0504031},  J. S. Alcaniz, Deepak Jain and Abha Dev, {\it
  Phys. Rev. D}, {\bf 67}, 043514 (2003).

\bibitem{jain}Deepak Jain, J. S. Alcaniz and Abha Dev, {\bf astro-ph/
  0409431} (To be published in Nuc. Phys. B).


 \bibitem{huterer}
 A. R. Cooray and D. Huterer, {\ApJ}, {\bf 513}, {L95} (1999).
 %
 %
\bibitem{efs}
G. Efstathiou, {\MNRAS}, {\bf 310}, 842 (1999).

\bibitem{linder}
M. Chevallier and D. Polarski, {\it Int. Jr. Mod. Phys.}, {\bf D10},
213 (2001);
E. V. Linder, {\it Phys. Rev. Lett}, {\bf 90}, 091301 (2003).

\bibitem{pad}
 T. Padmanabhan and T. Roy Choudhury, {\MNRAS}, {\bf 344}, 823 (2003). 

\bibitem{krat} J. Kratochvil, A. Linde, E. V. Linder and M. Shmakova,
  JCAP {\bf 0407}, 001 (2004). 

\bibitem{wsps} Y. Wang and P. M. Garnavich, {\ApJ} {\bf 552}, 445
  (2001); C. R. Watson and R. J. Scherrer, Phys. Rev. {\bf D68},
  123524 (2003); P.S. Corasaniti et al.,  
{\bf astro-ph/0406608}.

\bibitem{teg} Y. Wang and M. Tegmark, {\it Phys. Rev. Lett.} {\bf 92},
  241302-1 (2004). 

\bibitem{padn} H. K. Jassal, J. S. Bagla, and T. Padmanabhan,
  {\MNRAS}, {\bf 356}, L11 (2005)

\bibitem{padn1} H. K. Jassal, J. S. Bagla, and T. Padmanabhan,
  {\bf astro-ph/0506748} (2005).

\bibitem{Freedman}
W. L. Freedman et al.,{\ApJ}, {\bf 553}, 47 (2001).

\bibitem{KH}S. Komossa and G. Hassinger, in XEUS studying the
 evolution of the universe , G. Hassinger et al.(eds.), MPE report,
 in Press ( astro-ph/0207321)

\bibitem{HF}
F. Hamman and G. J. Ferland, {\ApJ}, {\bf 418} 11 (1993)

\bibitem{a}
 J. S. Alcaniz and J. A. S. Lima, {\ApJ}, {\bf550}, L133 (2001)

\bibitem{mc}
 P. J. McCarthy et al., {\ApJ}, {\bf614}, L9 (2004)

\bibitem{get}
Geetanjali Sethi, Abha Dev and  Deepak Jain, {\it Phys. Lett. B}, {\bf
  624}, 135 (2005)

\bibitem{v}U. Alam \& V. Sahni, {\bf astro-ph/ 0511473} (2005)

\bibitem{linear}
A. Batra, M.Sethi and D. Lohiya, {\it  Phys. Rev. D}, {\bf 60}, 108301
(1999); A. Dev et. al, {\it Phys. Lett. B}, {\bf548}, 12 (2002);
Deepak Jain, A. Dev and J. S. Alcaniz, {\it Classical \& Quantum
  Grav.}, {\bf 20}, 4163 (2003)


\bibitem{cap}
S. Capozziello et al., {\bf astro-ph/0410268}.

\bibitem{feng}
B. Feng, X. Wang and  X. Zhang, {\it Phys. Lett. B}, {\bf 607}, 35 ( 2005).


\bibitem{jim}
Raul Jimenez and A. Loeb, {\ApJ}, {\bf 573}, 37 (2002); Raul Jimenez
et al., {\ApJ}, {\bf 593}, 622 (2003). 

\bibitem{sim}
J. Simon, L. Verde and
Raul Jimenez, {\it Phys. Rev. D}, {\bf 71}, 123001 (2005).

\bibitem{st}P.B. Stetson, D.A.Vandenberg \& M. Bolte, {\it PASP}, {\bf
  108}, 560 (2001)
\end {thebibliography}
\end{document}